\documentclass{iaus}
\usepackage{graphicx}

\title[Imaging ``Pinwheels'' with interferometry] 
      {Imaging ``Pinwheel'' nebulae with optical long-baseline interferometry}
      
      \author[F. Millour et al.]   
             {F. Millour$^{1,2}$, T. Driebe$^{1,3}$, J. H. Groh$^1$,
               O. Chesneau$^2$, G. Weigelt$^1$, A. Liermann$^1$ \&
               A. Meilland$^1$}
             
             \affiliation{$^1$Max-Planck Institute for Radioastronomy, auf dem
               H\"ugel 69, 53121 Bonn, Germany
               \\[\affilskip]
               $^2$Observatoire de la c\^ote d'Azur, Bd de
               l'Observatoire, 06304 Nice, France
               \\[\affilskip]
               $^3$German Aerospace Center (DLR), K\"onigswinterer
               Str. 522-524, 53227 Bonn, Germany
               \\ Contact email: {\tt fmillour@oca.eu}}
             
             \pubyear{2010}
             \volume{xxx}  
             \pagerange{119--126}
             \jname{Active OB stars: Structure, Evolution, Mass loss
               and critical limits}
             \editors{C. Neinert, B.D. Editor \& C.E. Editor, eds.}
             
\begin{document}

             \maketitle

             \begin{abstract}
               Dusty Wolf-Rayet stars are few but remarkable in terms
               of dust production rates (up to $ \dot {\rm M} =
               10^{-6}{\rm M}_\odot$/yr). Infrared excesses
               associated to mass-loss are found in the sub-types
               WC8 and WC9. Few WC9d stars are
               hosting a ``pinwheel'' nebula, indirect
               evidence of a companion star around the primary. While
               few other WC9d stars have
               a dust shell which has been barely resolved so
               far, the available angular resolution offered by single
               telescopes is insufficient to confirm if they also host
               ``pinwheel'' nebulae or not. In this article, we
               present the possible detection of such
               nebula around the star WR\,118. We discuss about the
               potential of interferometry to image more
               ``pinwheel'' nebulae around other WC9d stars.
             \end{abstract}

             \firstsection 
             \section{WR\,118}
             
             In 2008, we observed the dusty Wolf-Rayet (WR) star
             WR\,118 using the Astronomical Beam Recombiner (AMBER,
             \cite{petrov}), at the focus of the Very Large Telescope
             Interferometer (VLTI). WR\,118 had already been observed
             using speckle interferometry, on the BTA 6m telescope, in
             Russia (\cite{2001A&A...379..229Y}). The AMBER
             visibilities and closure phases were acquired at spatial
             frequencies up to 5 times larger than the previous
             speckle observations.

             We clearly resolved the system with AMBER, with
             visibilities decreasing up to $\approx55$ cycles /
             arc-second, and increasing again above. Such a
             visibility behavior is typical of an object containing a
             sharp edge in its intensity distribution. In addition,
             the closure phase, measured at three different moments
             during the night, is clearly non-equal to zero, meaning
             that WR\,118's dusty nebula is asymmetric.
             
             We modelled WR\,118's dusty nebula with several
             geometrical models, including a clumpy spherical wind
             and a ``pinwheel'' nebula. This last model provides a
             physical description of the system and best match the
             observed data.

             Therefore, we concluded that WR\,118 probably hosts a
             ``pinwheel'' nebula, detected for the first time using
             long-baseline interferometry (\cite{Millour_etal09}).

             \section{Preparing future observations}

             Repeating the AMBER observations on WR\,118 to confirm
             its ``pinwheel'' nature is the next step in this research
             program. One step forward is to assess the feasibility of
             imaging such targets with the current and future
             capabilities of the VLTI.

             For that, we perform simulations of the current VLTI
             instrument AMBER (K-band, 2.2 microns) and of the second
             generation planned VLTI instrument MATISSE (L-band,
             3.5 microns) observations. We start from the model of
             WR\,118 found previously and simulate a
             realistic observation (typical V$^2$ errors of 5\%) with
             different baselines and one measurement per hour. Then, we
             use the MIRA software (\cite{2008SPIE.7013E..43T})
             to reconstruct an image of the target from these
             simulations. Four cases are simulated here:
             \begin{itemize}
             \item AMBER observations during 3 nights with 3
               telescopes configurations,
             \item AMBER observations during 7 nights with 7
               telescopes configurations,
             \item MATISSE observations during 3 nights with 3
               telescopes configurations,
             \item MATISSE observations during 7 nights with 7
               telescopes configurations.
             \end{itemize}

\begin{figure}[htbp]
               \begin{center}
                 \includegraphics[height=0.24\textwidth,
                   angle=-90]{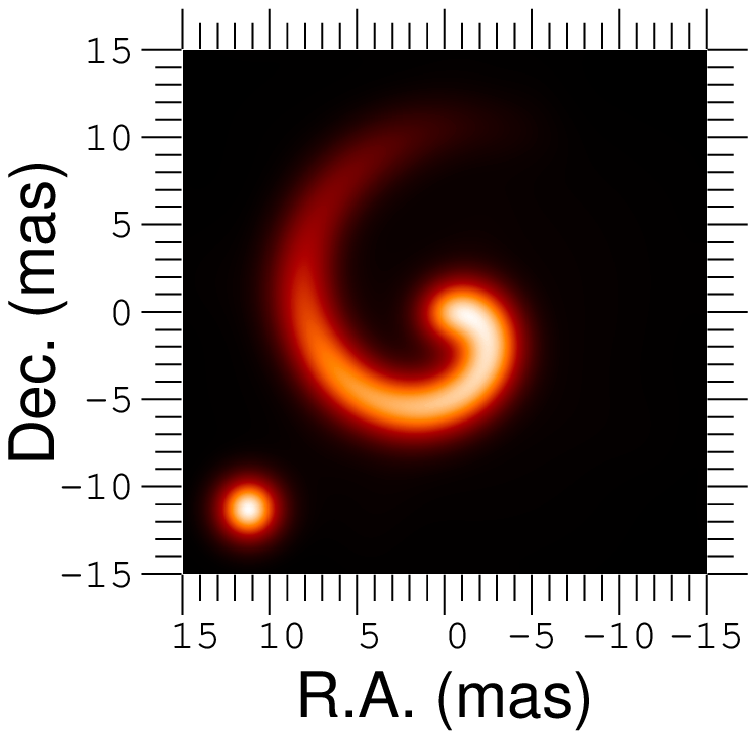}
                 \includegraphics[height=0.71\textwidth,
                   angle=-90]{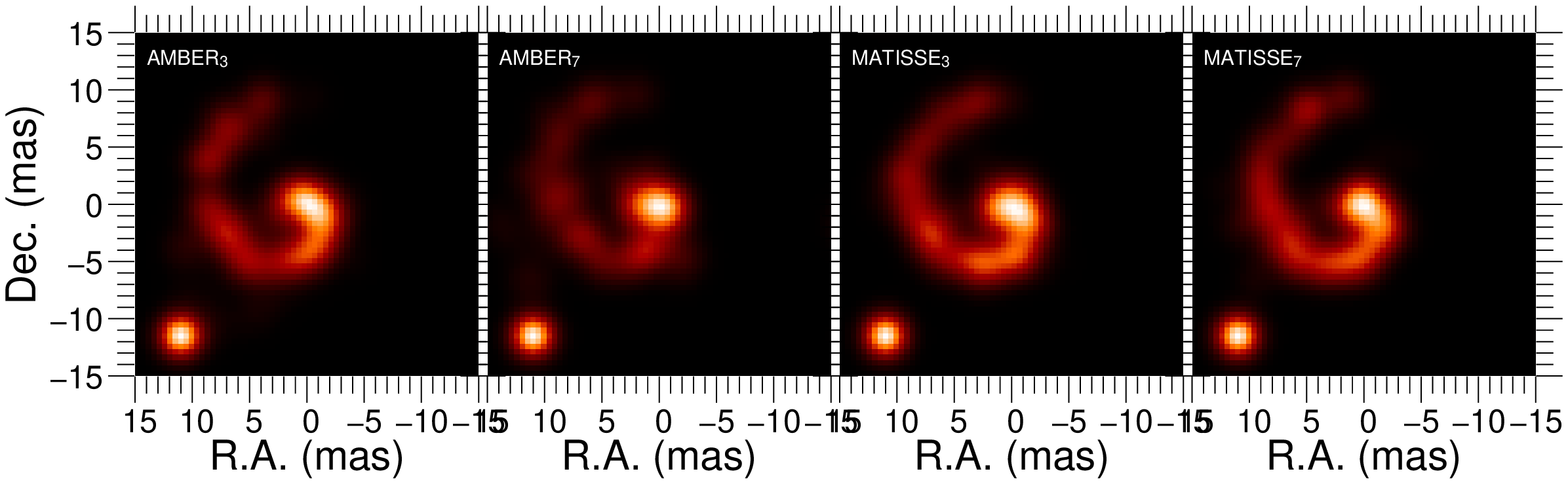}
                 \caption{Simulations of aperture synthesis of
                   ``pinwheel'' nebulae, using the parameters 
                   from WR\,118. From left to
                   right: the model used
                   convolved with a beam equivalent to a 130m diameter
                   telescope, synthesised images for AMBER
                   observations during 3 and 7 nights, and synthesised
                   images for MATISSE observations during 3 and 7
                   nights.}
                 \label{fig1}
               \end{center}
             \end{figure}

             As expected, we find that AMBER already provides some
             imaging capabilities, even using as few as three nights
             of observation. One can recognize in the reconstructed
             image all the features from the original image (see
             Fig.~\ref{fig1}). However, as was also experienced on
             real datasets (\cite{Millour2}), the relative flux of the
             different features is not well constrained, because of
             the lack of 66\% of phase information. Increasing, by
             more than a factor two, the number of nights used, only
             marginally improves the quality of the image
             reconstruction.

	     On the other hand, MATISSE simulations show a better
             agreement on the fluxes of the different features, even
             using ``only'' three nights of observation. We also note
             that the loss in angular resolution from the K-band  and
             L-band do not apparently affect the quality of the image
             reconstruction.

             \section{Conclusion}

             We showed here that long-baseline interferometry is now
             clearly mature to confirm the
             brightest suspected ``pinwheel'' nebulae, using the current
             abilities of AMBER on the VLTI. In addition, we also
             showed that AMBER will be able to synthethise an image,
             where the ``pinwheel'' nebula would be qualitatively
             recognized, using at least three nights of observing
             time. The use of the second generation MATISSE instrument
             will permit to have more realistic flux measurements
             in the different parts of the image, with the same amount
             of observing time.


\begin{thebibliography}{}


             \bibitem[Millour \etal\ 2009a]{Millour_etal09}
               {Millour, F., Driebe, T., Chesneau, \etal} 2009a,
               \textit{A\&A}, 506, 49

             \bibitem[Millour \etal\ 2009b]{Millour2}
               {Millour, F.; Chesneau, O.; Borges Fernandes, M.; \etal} 2009b,
               \textit{A\&A}, 507, 317

             \bibitem[Petrov \etal\ 2007]{petrov}
               {Petrov, R. G.; Malbet, F.; Weigelt, G.; \etal} 2007,
               \textit{A\&A}, 464, 1-12

             \bibitem[Thi\'ebaut 2008]{2008SPIE.7013E..43T}
               {Thi\'ebaut, E.} 2008, \textit{SPIE}, 7013, 43

             \bibitem[Yudin \etal\ 2001]{2001A&A...379..229Y}
               {Yudin, B.; Balega, Y.; Bl\"ocker, T.; \etal} 2001,
               \textit{A\&A}, 379, 229

             \end{thebibliography}
             \end{document}